\documentclass[twocolumn]{aa} 

\usepackage{graphicx}
\usepackage{txfonts}
\usepackage{caption}
\usepackage{subcaption}
\usepackage{rotating}
\usepackage[colorlinks=true,linkcolor=bluea,allcolors=blue]{hyperref}
\usepackage{xcolor}

\begin{document}

   \title{Scattered synchrotron emission and a giant torus revealed in polarized light in the nearest radio galaxy Centaurus~A\thanks{Based on observations made with ESO Very Large Telescope at the Paranal Observatory under programme ID~111.24ZY.003}}
   
   \titlerunning{Spectropolarimetry of Cen~A}

   \author{F. Marin\inst{1}          
          \and
          D. Hutsem\'ekers\inst{2} 
          \and        
          C.-Z. Jiang\inst{3,4}    
          \and        
          R. Antonucci\inst{5}         
          \and
          P. M. Ogle\inst{6}
          \and
          A. Bagul\inst{7}
          \and
          C. Ramos Almeida\inst{8,9}   
          \and
          M. Berton\inst{10}  
          }

   \institute{Universit\'e de Strasbourg, CNRS, Observatoire Astronomique de Strasbourg, UMR                 7550, 11 rue de l'universit\'e, 67000 Strasbourg, France\\
             \email{frederic.marin@astro.unistra.fr}
             \and            
                 Institut d’Astrophysique et de G\'eophysique, Universit\'e de Li\`ege, All\'ee du 6 Ao\^ut 19c, B5c, 4000 Li\`ege, Belgium
             \and 
                 CAS Key Laboratory for Research in Galaxies and Cosmology, Department of Astronomy, University of Science and Technology of China, Hefei, Anhui 230026, China
             \and
                 School of Astronomy and Space Science, University of Science and Technology of China, Hefei 230026, China
             \and                 
                 University of California, Physics Department, Santa Barbara, Broida Hall, Santa Barbara, CA 93106-9530, USA 
             \and  
                 Space Telescope Science Institute, 3700 San Martin Dr., Baltimore, MD 21218
             \and    
                 Centre for Extragalactic Astronomy, Department of Physics, Durham University, Durham DH1 3LE, UK
             \and            
                 Instituto de Astrof\'isica de Canarias, Calle V\'ia L\'actea, s/n, E-38205, La Laguna, Tenerife, Spain
             \and 
                 Departamento de Astrof\'isica, Universidad de La Laguna, E-38206, La Laguna, Tenerife, Spain
             \and            
                 European Southern Observatory (ESO), Alonso de C\'ordova 3107, Casilla 19, Santiago 19001, Chile  
             }

   \date{Received October 10, 2024; accepted February 7, 2025}
	
 
  \abstract
   {Centaurus A (Cen~A) is the closest radio galaxy and a prime example of a low-luminosity active galactic nucleus (AGN), exhibiting complex emissions across the electromagnetic spectrum. The nature of its continuum emission, particularly the mechanisms powering it, has been a subject of considerable debate due to the fact that the AGN is deeply buried in dust and therefore not directly observable.}
   {This study aims to elucidate the origin of the continuum emission in Cen~A and determine the geometrical arrangement of matter in the nuclear region by the mean of optical and near-infrared spectropolarimetry.}
   {We obtained spectropolarimetric data of Cen~A using the VLT/FORS2, covering the 6100 -- 10300~\AA\ spectral range with an effective resolving power of about 1000. Analysis was conducted on several regions near the obscured AGN, examining total and polarized fluxes, in order to find signatures of the AGN in scattered light.}
   {The analysis revealed a region showing strong and narrow emission lines associated with AGN activity. After correction for interstellar polarization in the dust lane (but not for starlight), the intrinsic polarization of the scattered AGN light exhibits a polarization degree of 2 -- 4\%, decreasing from optical to near-infrared, associated with a polarization position angle perpendicular to the radio jet axis. We exclude the presence of hidden broad line in our polarized flux spectrum at $\ge$ 99\% probability. Narrow emission lines are found to be strongly polarized and orthogonal to the jet position angle. We demonstrate that a beamed synchrotron jet, scattering onto the narrow line region (NLR) best fits all the observational properties reported in this paper and the literature. In this model, the base of the NLR is obscured by a giant ($\ge$ 10~pc) circumnuclear region and can only become visible through perpendicular scattering onto the outermost part of the NLR, naturally producing high polarization degrees and polarization angles perpendicular to the radio structure.}
   {This study provides strong evidence that Cen~A defines a new class of hidden-NLR AGNs in which two other objects naturally find their place (NGC~4258 and 3C~270) and supports old predictions that beamed synchrotron jets can be observed in reflection. Future surveys should focus on identifying similar hidden-NLR AGNs, especially among misdirected BL~Lac AGNs.}


   \maketitle
%

\section{Introduction}
\label{Introduction}

Centaurus A (Cen~A, NGC~5128, Caldwell~77) is the nearest radio galaxy from Earth, located at a distance of approximately 3.8~Mpc \citep{Harris2010}. At this scale, 1" represents 18.4~pc, which allows the distribution of gas, dust, and electrons to be probed in detail at very good spatial resolutions \citep[e.g.][]{Kraft2000,Radomski2008,Ramos2009,Janssen2021}. Its proximity makes it a valuable subject for studying the detailed mechanisms of active galactic nuclei (AGN). Indeed, Cen~A is a Fanaroff-Riley type I (FR~I) radio galaxy, exhibiting a combination of characteristics found in either Seyfert or radio galaxies : powerful, extended emission at radio frequencies \citep{Burns1983}, high-energy X-ray and $\gamma$-ray emission from its compact core \citep{Kraft2000,HESS2018}, low bolometric luminosity ($\sim$ 10$^{43}$ erg~s$^{-1}$, \citealt{Borkar2021}) and shock-excited low-ionization emission lines in the nuclear disk \citep{Simpson1998}. See \citet{Antonucci2012} for a review on the nature of FR~I radio galaxies, among other AGN classes.

Among the other features of this source, not caused by the AGN but rather to the history of galaxy mergers, is a prominent kiloparsec-scale dust lane, which bisects the galaxy and obscures the central AGN \citep{Schreier1996}. This dust lane poses a challenge for studying the nucleus at ultraviolet, optical, and near-infrared wavelengths, where distinctive features of the accretion/ejection mechanisms powered by a supermassive black hole leave their imprints. Indeed, the nature of the continuum emission in Cen A has been widely debated, with proposed mechanisms including synchrotron radiation from relativistic jets and thermal emission from an accretion flow \citep[e.g.][]{Bailey1986,Packham1996,Whysong2004,Meisenheimer2007}. The underlying question is: what is the source of FR~I radio galaxies emission? Thermal emission and/or synchrotron emission?

Total-intensity light photometry, spectroscopy, imaging, and timing analyses have proven to be ineffective in the case of Cen~A due to the excess of obscuring material, and it is not an isolated case : there are numerous radio galaxies where the entire nuclear region is obscured by a kiloparsec-scale dust lane \citep{Antonucci1990}. Yet, polarimetry can take a unique look at this problem since it is able to detect light from the obscured core that has scattered on distant material, therefore offering us a periscopic view of the system. Attempts were made in the infrared to detect the presence of a broad Br$\gamma$ line in polarized flux but, due to insufficient photons count, it was inconclusive \citep{Alexander1999}. In fact, all the polarimetric measurements made in the near and mid-infrared were inconclusive to determine if the observed nuclear polarization is owing to scattered light from the nucleus or if it originates from synchrotron emission from a BL Lac source inclined to our line-of-sight (see \citealt{Marin2023} for a review). The main reason for this is that all the polarimetric measurements (either broadband polarimetry or spectropolarimetry) were done blindly, that is to say by targeting the brightest source at the heart of Cen~A, without being certain that this corresponded to a scattering region and not to a transmission region. This was proven by \citet{Schreier1996} who first obtained a high-resolution HST polarization map of the nuclear region of Cen~A and discovered that the peak of polarization was not centered on the supposed location of the nucleus, nor along the radio jet position angle, but several tens of parsecs away from the infrared flux peak. Unfortunately, since then, no one has observed this particularly promising subarcsecond region using spectropolarimetric techniques. This was the goal of the spectropolarimetric observations we achieved and which we present in this article. 

The paper is structured as follows: Sect.~\ref{Observation} describes the observations and data reduction. Sect.~\ref{Analysis} presents the results of the spectropolarimetric analysis, focusing on three different regions near the obscured core of Cen~A. Sect.~\ref{Discussion} discusses the implications of our findings for the nature of the continuum emission in Cen~A but also for the geometric arrangement of matter around its central supermassive black hole. Finally, Sect.~\ref{Conclusion} summarizes our conclusions and outlines potential directions for future research.

\section{Observations and data reduction}
\label{Observation}

The initial target of this observing campaign was the high-polarization ($>$ 8\%) compact emission knot found by \citet{Schreier1996} at $\sim$ 1.4" (26~pc) from the supposed location of the hidden AGN, bluer than the surrounding emission. According to their work, the knot is "bluer or less obscured than the surrounding emission", but the kilo-parsec dust lane is on a much larger scale than the region we observed. The reddening due to the kpc-scale dust is certainly shared by the nucleus. It means that the polarized knot is intrinsically bluer, rather than less obscured. This makes even more sense when accounting for the presence of the AGN, as revealed by the emission lines, since AGN nuclei are known to be bluer that the bulge stars \citep{Bruce2016,McPartland2019}. We will report structured obscuration on nuclear scales as well later in this work.

\begin{figure}
    \centering
    \includegraphics[trim={0cm 0cm 0cm 0cm}, clip, width=\linewidth]{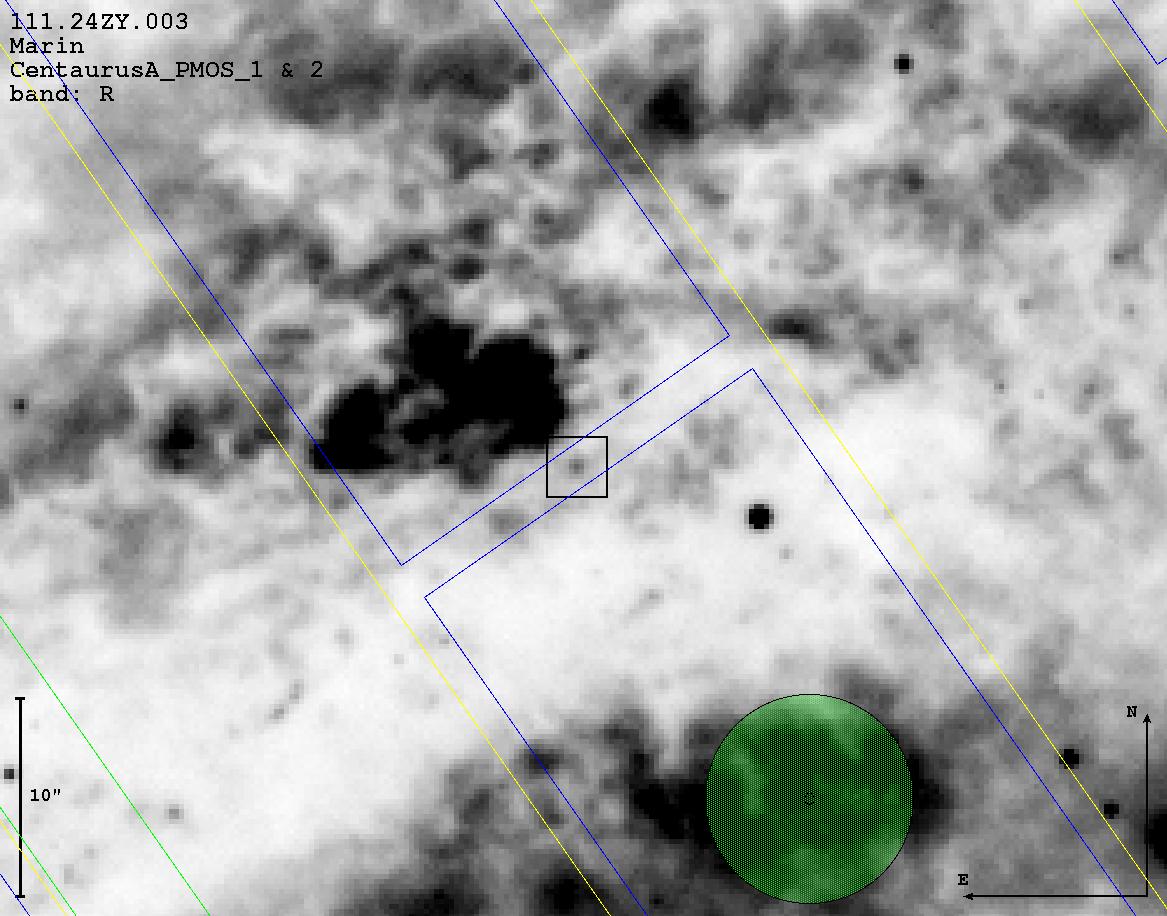}
    \caption{Position and orientation of the MOS slit superimposed on the Centaurus A galaxy dark lane seen in R band. The 2"-wide slit is formed by positioning two opposite MOS blades (shown in blue). It is centered on the polarized knot (indicated by a square). The scale and orientation of the field are indicated. The green disk is the FORS Instrument Mask Simulator (FIMS) control button.}
    \label{fig:slit}
\end{figure}

To probe this polarized knot with greater details, we first secured a good-seeing (0.5") R-band image of the Cen~A region with the FOcal Reducer/low dispersion Spectrograph 2 (FORS2) mounted on the Very Large Telescope (VLT) at the Paranal Observatory on February 19/20, 2023 (ObsID : 3506985) to precisely put the slit on the "polarized knot". We also observed the Cen~A region in the vicinity of the knot with imaging polarimetry in the R-band to accurately characterize the polarization in the surroundings on June 14/15, 2023 (ObsID : 3507092). For this observation, the field was rotated 35\degr\ clockwise. Then, we proceeded with actual spectropolarimetry on June 17/18, 2023 (Obs ID : 3571662) and June 19/20, 2023 (Obs ID : 3581876) using the grism 300I + OG590 (four positions of the half-wave plate (HWP) were used in each observation block: 0, 22.5, 45, and 67.5\degr). The 2" slit was positioned roughly along the dark lane, with a position angle of 125\degr (see Fig.~\ref{fig:slit}). The pixel size was 0.25" $\times$ 0.25" on the sky. Our goal was to examine the H$\alpha$ line and as much of the optical-near-infrared continuum as possible. The wavelength range covered by the grism is 6000 -- 11000 \AA , and the resolving power is R = 660 for a 1" slit, but in this case, the final spectral resolution is seeing- and not slit-dominated. Observational details are listed in Tab.~\ref{tab:observation}.

\begin{table*}
    \caption{Observation log.}     
    \label{tab:observation}      
    \centering                                      
    \begin{tabular}{c c c c c c c c }        
    \hline\hline              
    \textbf{Date} & \textbf{ObsID} & \textbf{Purpose} & \textbf{Observing time} & \textbf{Seeing} & \textbf{Sky Transparency} & \textbf{Airmass} & \textbf{FLI} \\  
    ~ & ~ & ~ & \textbf{(hh:mm:ss)} & \textbf{(")} & ~ & ~ & ~ \\  
    \hline                        
     19/20-Feb-2023 & 3506985 & Pre-imaging	& 00:05:46	& 0.5 & clear & 1.1 & 0.0 \\\relax
     14/15-Jun-2023 & 3507092 & Imaging polarimetry & 00:22:23 & 1.0 & clear & 1.1 & 0.1 \\\relax
     17/18-Jun-2023 & 3571662 & Spectropolarimetry & 01:20:24 & 0.8 & clear & 1.3 & 0.0 \\\relax
     19/20-Jun-2023 & 3581876 & Spectropolarimetry & 01:17:09 & 0.6 & clear & 1.1 & 0.0 \\ 
    \hline                                          
    \end{tabular}
    \tablefoot{FLI means "fraction of lunar illumination".}
\end{table*}

\begin{figure*}
    \centering
    \includegraphics[trim={0cm 0cm 0cm 0cm}, clip, width=\linewidth]{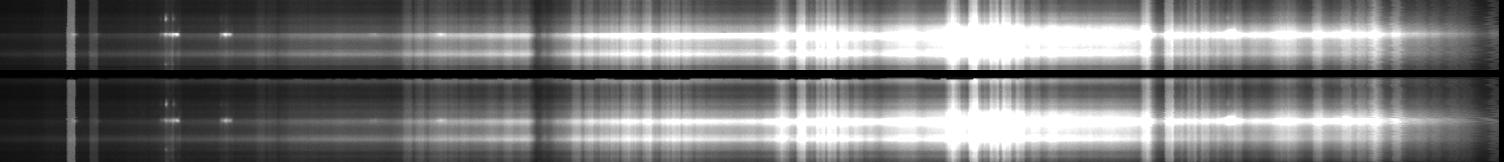}
    \caption{2D spectrum (both ordinary and extraordinary rays) obtained by the VLT/FORS2 of the polarized knot (top) and SE cloud (bottom). The wavelength range is 6100 -- 10300 \AA, short wavelengths to the left. The displayed slit height is around 20".}
    \label{fig:2D_spectrum}
\end{figure*}

\begin{figure}
    \centering
    \includegraphics[trim={0cm 0cm 0cm 0cm}, clip, width=\linewidth]{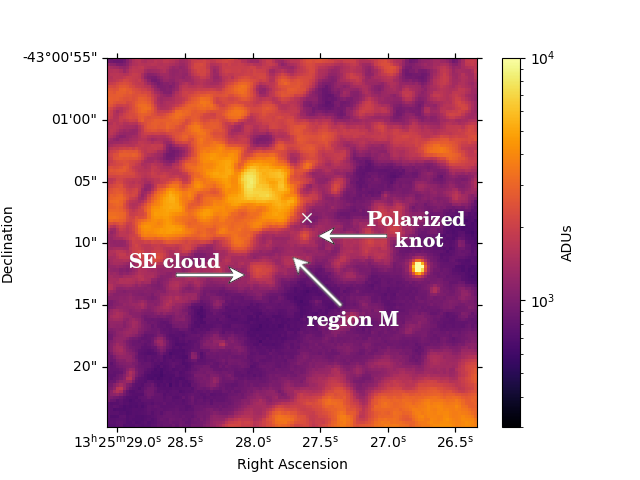}
    \caption{R-band map of the central region of Cen A, in Analog to Digital Units (ADUs). The right arrow points towards the polarized knot discovered by \citet{Schreier1996} and the left arrow points towards a South-East (SE) cloud that we had the curiosity to examine. The region in-between (labeled as M) is also studied in this paper, as it is likely completely dominated by starlight and can therefore inform us about the impact of dichroic absorption on the polarimetric signal. The white cross marks the supposed location of the supermassive black hole, established from radio measurements \citep{Ojha2010}. The bright dot on the East is a foreground star (Gaia DR3 6088704625623266048). North is up, East is left.}
    \label{fig:ADU}
\end{figure}

Raw frames were first processed to remove cosmic ray hits using the python implementation of the ``lacosmic'' package \citep{2001VanDokkum,2012VanDokkum}. The FORS2 pipeline \citep{2019vltman} was then used to get images with two-dimensional spectra rectified and calibrated in wavelength. Though VLT pipeline provides an automated source-detect-and-extract algorithm from a 2-dimensional (2D) spectrum to a 1-dimensional (1D) spectrum, we deem it necessary to start our reduction with the 2D spectrum product to identify the regions of interest because the flux of our target is both diffuse and embedded in a strong background. From the 2D spectrum, shown in Fig.~\ref{fig:2D_spectrum}, we were surprised to discover, in addition to Schreier's knot, a second local peak of flux inside our slit. Located South-East of the polarized knot, we will thus label it as the "SE cloud" henceforth. 

Due to the dual-beam design on the VLT FORS, the o-ray (ordinary) and e-ray (extraordinary) are recorded simultaneously and projected onto different positions on the detector with an offset in spatial direction (88 pixels or 22"). The subsequent reduction is performed for the o-ray, including source extraction and sky subtraction. The same reduction for the e-ray was accomplished by adjusting the windows based on the offset.

We employed 2.75"-long and 2.0"-wide apertures (corresponding to 11 pixels in length at a resolution of 0.25"/pixel) to extract the 1D spectrum of both the SE cloud and the polarized knot, centered on the peaks of flux in the spatial direction. A smaller 1.5"-long and 2.0"-wide aperture was used to extract the total and polarized fluxes from the region in between the knot and the SE cloud, where obscuration seems to be predominant, allowing us to have a background reference spectrum within Cen~A. We label this region as "M" for simplicity. The spatial position of each of the extracted regions is indicated in Fig.~\ref{fig:ADU}. 

All the regions of interest mentioned in the previous paragraph are on the same slit, allowing for a uniform sky subtraction for all three regions. The sky spectrum was obtained using a MOS slit located 2.7' upward on the CCD frame, where there is little galaxy light remaining. The same procedure was repeated for all four position angles of the polarizers (0$^{\circ}$, 22.5$^{\circ}$, 45$^{\circ}$, 67.5$^{\circ}$), resulting in 8 spectra. The sky-subtracted o-ray and e-ray were then combined to produce normalized Stokes parameters $q$ and $u$ using the following formulas:
\begin{equation}
q = Q/I = \frac{1}{2}(\frac{f_{o,0}-f_{e,0}}{f_{o,0}+f_{e,0}} -\frac{f_{o,45}-f_{e,45}}{f_{o,45}+f_{e,45}})
\end{equation}
\begin{equation}
u = U/I = \frac{1}{2}(\frac{f_{o,22.5}-f_{e,22.5}}{f_{o,22.5}+f_{e,22.5}} -\frac{f_{o,67.5}-f_{e,67.5}}{f_{o,67.5}+f_{e,67.5}})
\end{equation}
where $f_{r,\alpha}$ is the intensity measured in the aperture for $r=o,e$, $\alpha$ is the HWP position angle, and $I = (f_{o,0}+f_{e,0}+f_{o,45}+f_{e,45}+f_{o,22.5}+f_{e,22.5}+f_{o,67.5}+f_{e,67.5})/4$.

The usual polarization quantities $P$ (linear polarization degree) and $\theta$ (electric vector position angle) are computed from $q$ and $u$ using the standard equations :
\begin{equation}
P = \sqrt{q^{2}+u^{2}}
\end{equation}

and 
\begin{equation}
\theta = \frac{1}{2}\arctan\,(\frac{u}{q}).
\end{equation}

Finally, because the foreground material in our Galaxy is contributing no more than 0.5\% to the observed polarization \citep{Hough1987}, we neglected the impact of Galactic interstellar polarization on our final spectra. The polarization arising from dichroic interstellar absorption in Cen A is, however, very strong \citep{Bailey1986,Schreier1996} and should be accounted for. This is the reason why we extracted the signal from region M, where interstellar polarization is thought to dominate. We will explore this further in the next chapter.

\section{Analysis}
\label{Analysis}

The approximate position of the supermassive black hole powering Cen~A jets is 13h25m27.s6152, -43$^\circ$01'08.''805 \citep{Ojha2010,Janssen2021}. At those coordinates (reported in Fig.~\ref{fig:ADU}, but see Sect.~\ref{Discussion:hNLR} for slightly different values), the amount of dust obscuring the AGN core is so high that only starlight and dichroic absorption signatures from the dust lanes are detected in the optical and near-infrared bands. In hopes to glimpse scattered AGN signatures, we must look into the surroundings of the putative position of the supermassive black hole. We listed in the previous section three interesting structures that we will explore in great detail in the following. Those structures are 1) the polarized knot discovered by \citet{Schreier1996}, 2) a cloud South-East to the polarized knot that shows a strong continuum and no emission lines, and 3) the region M between positions 1 and 2 that is highly obscured and whose emission should be thus entirely dominated by starlight affected by interstellar polarization.

\subsection{The Schreier's polarized knot}
\label{Analysis:knot}

\begin{figure*}
    \centering
    \includegraphics[trim={0cm 0cm 0cm 0cm}, clip, width=\linewidth]{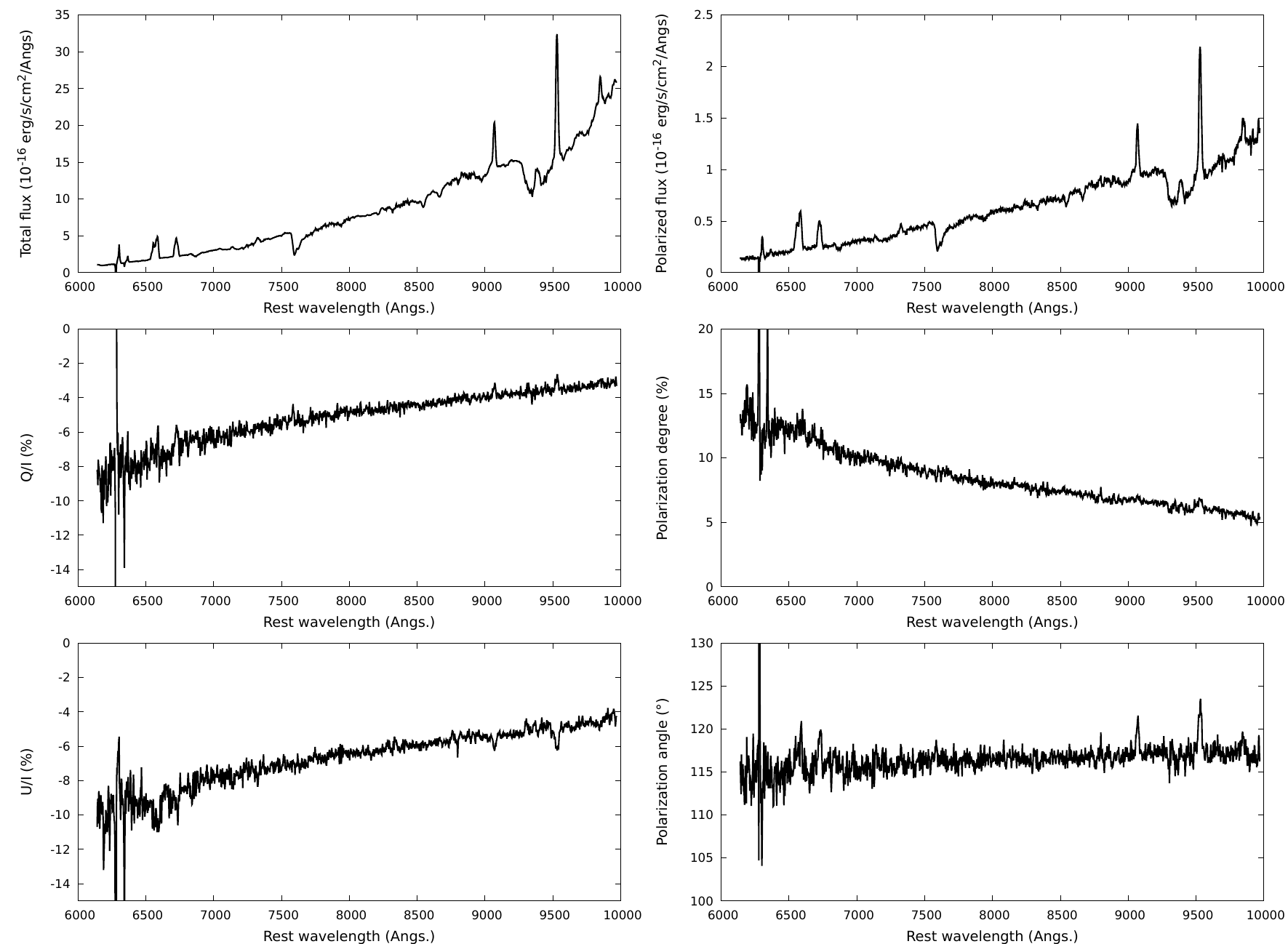}
    \caption{VLT/FORS2 spectropolarimetry of Centaurus~A's polarized knot. The total flux spectrum is shown in the top-left panel. The Stokes parameters Q and U (normalized by I) are shown on the middle-left and bottom-left panels, respectively. The polarized flux (that is the multiplication of the total flux with the polarization fraction) is shown on the top-right left panel. Finally, the polarization degree $P$ is presented in the middle-right panel and the polarization position angle $\theta$ is shown in the bottom-right panel.}
    \label{fig:Polarized_knot}
\end{figure*}

We begin our investigation by extracting the total and polarized flux spectra of the polarized knot discovered by \citet{Schreier1996}, as it is thought to be the best place to look for scattered light from the hidden nucleus. The 6152 -- 9988~\AA\, rest-frame spectra of the polarized knot are presented in Fig.~\ref{fig:Polarized_knot}. 

Focusing on the total flux spectrum first, we observe a continuum that is heavily reddened, something that can be trivially ascribed to extinction by the dust lane, with 22 times more flux at 9500~\AA\, than at 6170~\AA. On top of the continuum are absorption lines from two different origins. First, and most visible, is telluric contamination by the Earth's atmosphere, revealed by O$_2$ and H$_2$O absorption lines (see, e.g., \citet{Smette2015} and \citet{Koss2022}). Second, are absorption lines at 8498~\AA\,, 8542~\AA\, and 8662~\AA\, which belong to the well-known calcium triplet, i.e. three ionized calcium spectral lines that are most prominently observed in the absorption of spectral type G, K and M stars. Such lines indicate that our spectrum is contaminated by host starlight, a piece of information that will prove crucial for the rest of this analysis. Finally, and most importantly, we detect emission lines. 

\begin{figure}
    \centering
    \includegraphics[trim={0cm 0cm 0cm 0cm}, clip, width=\linewidth]{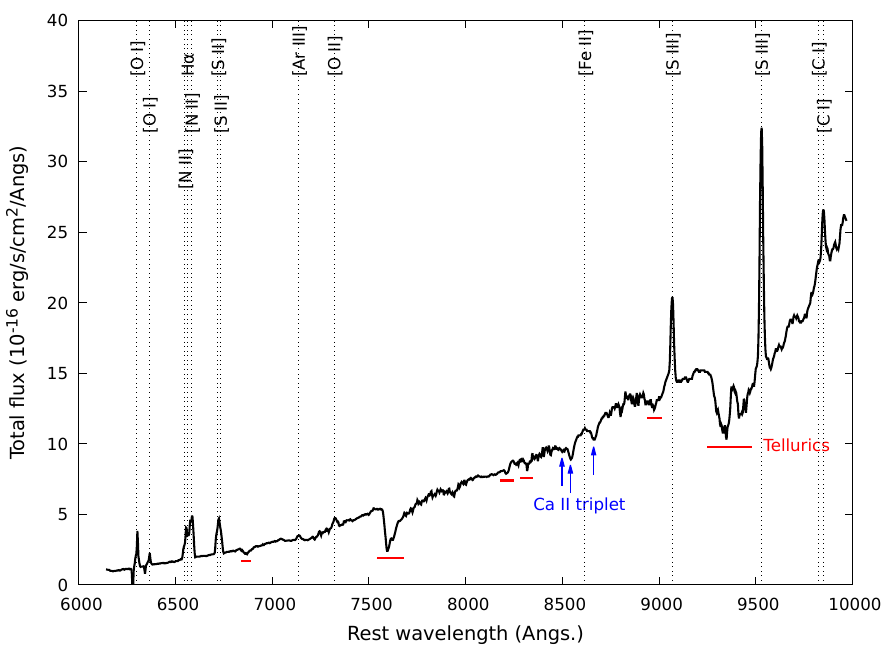}
    \caption{Tellurics (absorption), host starlight (absorption) and AGN (emission) lines detected in the total flux spectrum of the polarized knot. The optical and near-infrared AGN emission lines are produced by ionized gas and labeled in black. The telluric lines (in red) are from the Earth atmosphere. The stellar absorption lines (the calcium triplet) are indicated in blue.}
    \label{fig:lines}
\end{figure}

The presence of strong emission lines is a clear indicator of the presence of an AGN core. We indicate in Fig.~\ref{fig:lines} which absorption and emission lines correspond to which elements. Although the emission lines in the knot spectrum are only partially resolved\footnote{The slit was 2" wide but the resolution is fixed by the seeing at the time of the observation (0.6"-0.8"). For the grism used for this observation, the effective resolving power is then around 1000 (that is  $\sim$ 300~km~s$^{-1}$ full width at half maximum).}, it is evident that there are no broad (full width at half-maximum larger than $\sim$ 1000~km~s$^{-1}$) lines observed in the total flux spectrum of the polarized knot. In the frame of the unified model of AGN, it indicates that either the AGN core is obscured by dust and/or that there is no broad line emission at all. In passing, we notice that the polarized knot spectrum shows strong similarities to Low-Ionization Nuclear Emission-line Region (LINER) spectra, where low-ionization lines\footnote{We note the presence of high ionization lines such as [Si IV] and [Ne V] at position angles near the radio jet axis in the mid-infrared spectrum of Cen~A, possibly excited by ionizing radiation escaping from the AGN core \citep{Quillen2008}.} ([N~II], [S~II], [O~II] and [O~I]) are all relatively prominent \citep{Peterson1997}. The resemblance to AGN spectra is further reinforced by the intensity ratio of less reddened lines, such as [S~III] $\lambda_{\rm 9531}$/$\lambda_{\rm 9069}$ for which we find a value of 2.8, that is within the weighted average observed value of 2.55 $\pm$ 0.5 reported by \citet{Mendoza1983} and \citet{Osterbrock1990}.

Now, looking at the polarized signatures of the knot, we see in Fig.~\ref{fig:Polarized_knot} that the Q/I and U/I spectra, as well as the polarization degree $P$ spectrum, show a strong and almost featureless wavelength-dependent polarization (the strong artifacts seen in Q/I, U/I, $P$ and $\theta$ around 6300~\AA\, are due to absorption features that bring the fluxes down to zero). The linear polarization degree ranges from $\sim$ 15\% in the optical to $\sim$ 5\% in the near-infrared. The wavelength dependence of $P$ indicates that dust interactions (either scattering and/or dichroic absorption) prevails over electron scattering (Thomson scattering being wavelength-independent), regardless of the origin of the continuum (thermal emission from the accretion flow or synchrotron emission from the jet). The polarization position angle rotates slowly with increasing wavelengths (from 112$^\circ$ in the optical to 118$^\circ$ in the near-infrared). The fact that the near-infrared polarization angle evolves towards the value observed by \citet{Bailey1986} at 2 microns (147$^\circ$) and identified as being due to scattering strengthens the hypothesis that the intrinsic polarization of the knot comes from scattering off one single (extended) region, likely the NLR revealed by the presence of strong and narrow emission lines. In fact, the polarization angle in the narrow lines seems to deviate from that of the dust lane, reinforcing this hypothesis.  

To check this, we examine the polarized flux from the knot, that is the multiplication of the total flux with the polarization degree. We find that the resulting polarized spectrum is not only much bluer than the total flux (another tell-tale signature of dust scattering, see \citealt{Goodrich1994}) but also that there are no broad lines in polarized flux. In most type-2 AGNs, the region responsible for the emission of the broad lines (the BLR) is obscured by the circumnuclear material and can only be revealed in polarized light, as BLR photons perpendicularly scatter off the NLR into our line-of--sight (see the multiple examples in \citealt{Antonucci1985}, \citealt{Miller1990}, \citealt{Young1995}, \citealt{Tran1995}, \citealt{Ramos2016}). The absence of broad lines in the polarized flux of the knot indicates that there is likely no BLR, and that the light scattered by the knot is either originating from the accretion flow or from the jet itself.

In addition, we notice that the narrow emission lines do stand out in polarized flux ($PF_{\rm line} > PF_{\rm cont}$), indicating that they are polarized. Normally, photoionization and collisional excitation processes produce narrow emission lines that are little-to-no polarized. To check whether the polarization in the narrow lines is similar to or different from that of the continuum, we measured their polarization degree and angle, both in the line and in the blue and red continuum around the line. We report the values in Tab.~\ref{tab:line_polarization} and find that the lines and the continuum share the same polarization properties. It means that both the continuum and the narrow emission lines are likely being scattered by the same (dusty) medium. In the process, we do not find the very strong dependence of line polarization on critical density (the higher the critical density, the larger $P$), like in NGC~4258 \citep{Barth1999}, suggesting that both the continuum and lines polarization we observe is not entirely intrinsic to the source and must be affected by a second polarization source.

At this point, we have discovered that the polarized knot scatters the continuum emission from the obscured AGN. A large fraction of the polarization degree can be attributed to dust scattering, very likely happening in the NLR. We detected no broad emission lines in the polarized spectrum, indicating that either there is no BLR or that the knot is scattering the beamed synchrotron emission from the jet itself. However, two details do not fit the frame. First, the polarization angle is not exactly perpendicular to the direction of the jets, whose position angle was measured to be 55$^\circ$ $\pm$ 7$^\circ$ \citep{Burns1983}. We would expect a value close to 145$^\circ$, as it was measured by \citet{Bailey1986}, who found 147$^\circ$ at 2 microns, where dust extinction is less significant. The value we obtain is, in fact, closer to that of the dust lane (110$^\circ$ -- 115$^\circ$, \citealt{Bailey1986}). It indicates that part of the polarization we measured is a blended effect of nuclear polarization and starlight emission interacting with the dust lane, leading to dichroic absorption and thus non-null values of $P$ and $\theta$ for the interstellar polarization component. This would explain the second finding in tension with our discovery, that is to say, the fact that the polarization degree we measured in the near-infrared is only 5\%, while the 2 microns measurement obtained by \citet{Bailey1986} is as high as 9\% (after correcting for the polarization caused by the dust lane and host starlight). Our optical and near-infrared data are polluted. To confirm that the knot is scattering the AGN light and to measure the intrinsic polarization of the source, it is then crucial to correct the blended signal for interstellar polarization (that contains the underlying continuum).

\begin{table}
    \caption{Measurement of the knot polarization.}     
    \label{tab:line_polarization}      
    \centering                                      
    \begin{tabular}{c c c c c}        
    \hline\hline              
    Element & Q/I & U/I & $P$ (\%) & $\theta$ ($^\circ$) \\  
    \hline                        
     H$\alpha$+[N~II]$_{\rm cb}$   	&	-0.079	&	-0.093	&	12.25	&	  114.8 \\\relax
     H$\alpha$+[N~II]      	&	-0.073	&	-0.101	&	12.44	&	  117.1 \\\relax
     H$\alpha$+[N~II]$_{\rm cr}$   	&	-0.077	&	-0.096	&	12.29	&	  115.6 \\\relax
     [S~II]$_{\rm cb}$      	&	-0.075	&	-0.092	&	11.87	&	  115.3 \\\relax
     [S~II]         	&	-0.058	&	-0.093	&	10.99	&	  119.1 \\\relax
     [S~II]$_{\rm cr}$      	&	-0.067	&	-0.086	&	10.90	&	  115.9 \\\relax
     [S~III]$_{\rm cb}$     	&	-0.040	&	-0.056	&	6.91	&	  117.2 \\\relax
     [S~III]        	&	-0.033	&	-0.061	&	6.97	&	  120.8 \\\relax
     [S~III]$_{\rm cr}$    	&	-0.040	&	-0.053	&	6.61	&	  116.6 \\\relax
     [S~III]$_{\rm cb}$     	&	-0.036	&	-0.052	&	6.29	&	  117.8 \\\relax
     [S~III]        	&	-0.029	&	-0.061	&	6.75	&	  122.3 \\\relax
     [S~III]$_{\rm cr}$     	&	-0.034	&	-0.050	&	6.03	&	  117.7 \\
    \hline                                          
    \end{tabular}
    \tablefoot{Measurement in the brightest emission lines and in the adjacent 5 spectral bins continuum (cb =  blue continuum, cr = red continuum). Errors are about 0.4\% on $P$ and 1$^\circ$ on $\theta$.}
\end{table}

\subsection{The SE cloud}
\label{Analysis:SE_cloud}

\begin{figure*}
    \centering
    \includegraphics[trim={0cm 0cm 0cm 0cm}, clip, width=\linewidth]{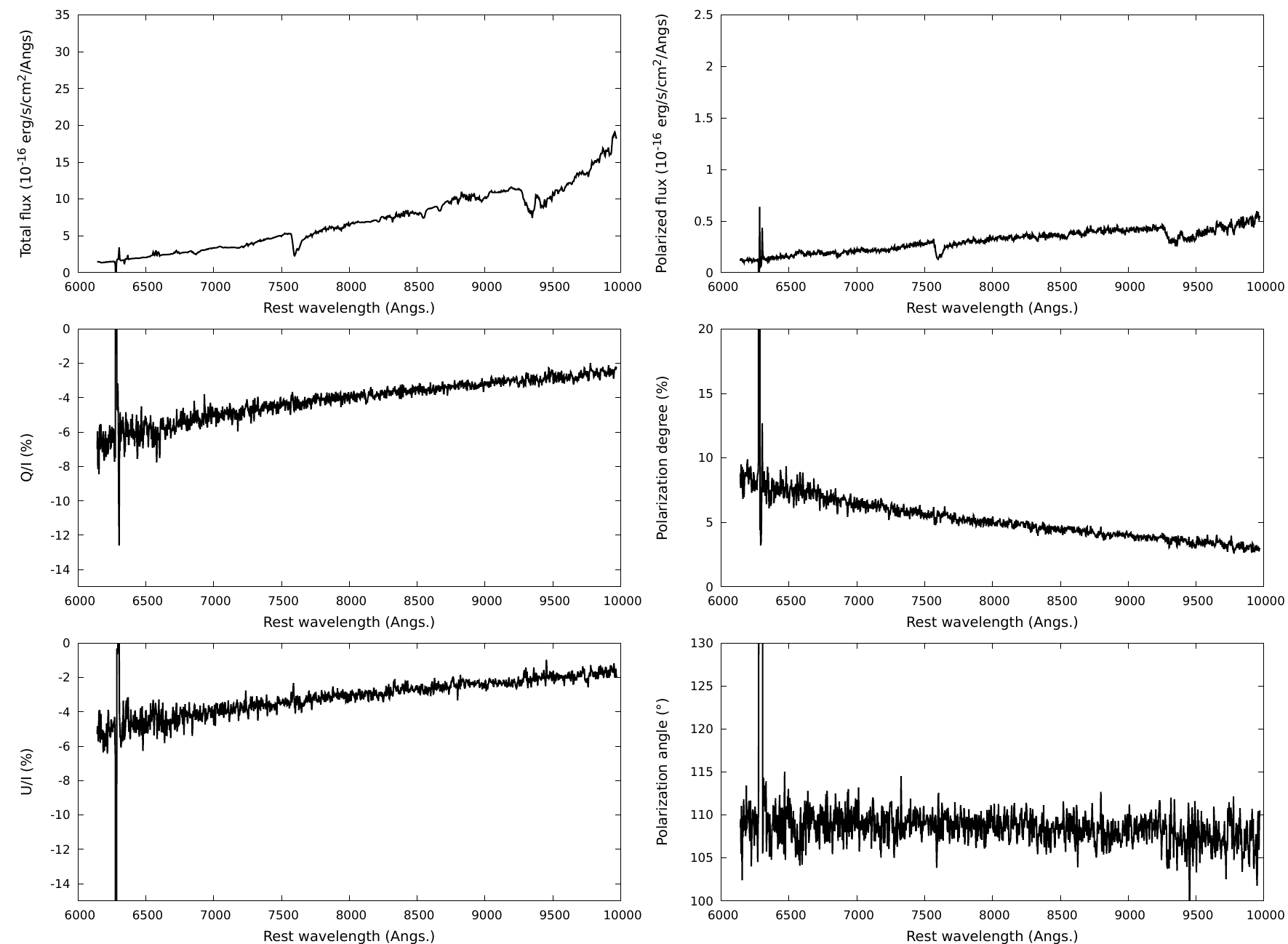}
    \caption{Same as Fig.~\ref{fig:Polarized_knot} but for the SE cloud identified in the 2D spectrum presented in Fig.~\ref{fig:2D_spectrum} and pinpointed in Fig.~\ref{fig:ADU}. The $y$ axes share the same scales as in Fig.~\ref{fig:Polarized_knot} to facilitate comparisons.}
    \label{fig:SE_cloud}
\end{figure*}

We now examine the SE cloud, which is the surprisingly bright region found in the 2D spectra of our observation, a small diffuse region 4.5" (83~pc in projected distance) south-east of the polarized knot. The total and polarized flux spectra are presented in Fig.~\ref{fig:SE_cloud}. From the total flux, we can see that the spectrum of the SE cloud consists only of a heavily reddened continuum and absorption features, partly from telluric and partly from stellar absorption (the calcium triplet). The absence of emission lines indicates that this cloud is not bright because it scatters light from the AGN, but because it likely suffers from less foreground extinction. This is coherent with the fact that, compared to the knot, the spectrum is slightly bluer (less absorbed than the knot). It is not an instrumental effect, since we can see on the 2D spectrum (Fig.~\ref{fig:2D_spectrum}) that the SE region is as intense as the knot in the blue part (left) but becomes much weaker in the red (right, by a factor $\sim$ 1.76 at 6500~\AA). The data reduction was identical for the two spectra, which are spatially very close. With a 2" slit and 0.8" seeing, it is neither due to seeing variation. It means that the knot is more reddened than the SE cloud.

Such a result is confirmed by looking at the polarization degree, which has the same trend as the polarized knot (i.e. $P$ decreases from the optical to the near-infrared) but $P$ is lower by several percent and its spectral slope is also slightly different from that of the knot (a $P$ $\sim$ $\lambda^k$ fit yields k = -1.9 in the case of the polarized knot and k = -2.1 for the SE cloud). The SE cloud polarization angle, almost wavelength-independent, shares the same polarization angle (109$^\circ$ on average) as that of the dust lane. It is thus clear that the SE cloud polarization results directly from dust dichroic absorption, without any detectable AGN contribution. We can conclude that the SE cloud is just a patch of pure starlight with lower foreground extinction.

\subsection{The region M}
\label{Analysis:region_M}

\begin{figure*}
    \centering
    \includegraphics[trim={0cm 0cm 0cm 0cm}, clip, width=\linewidth]{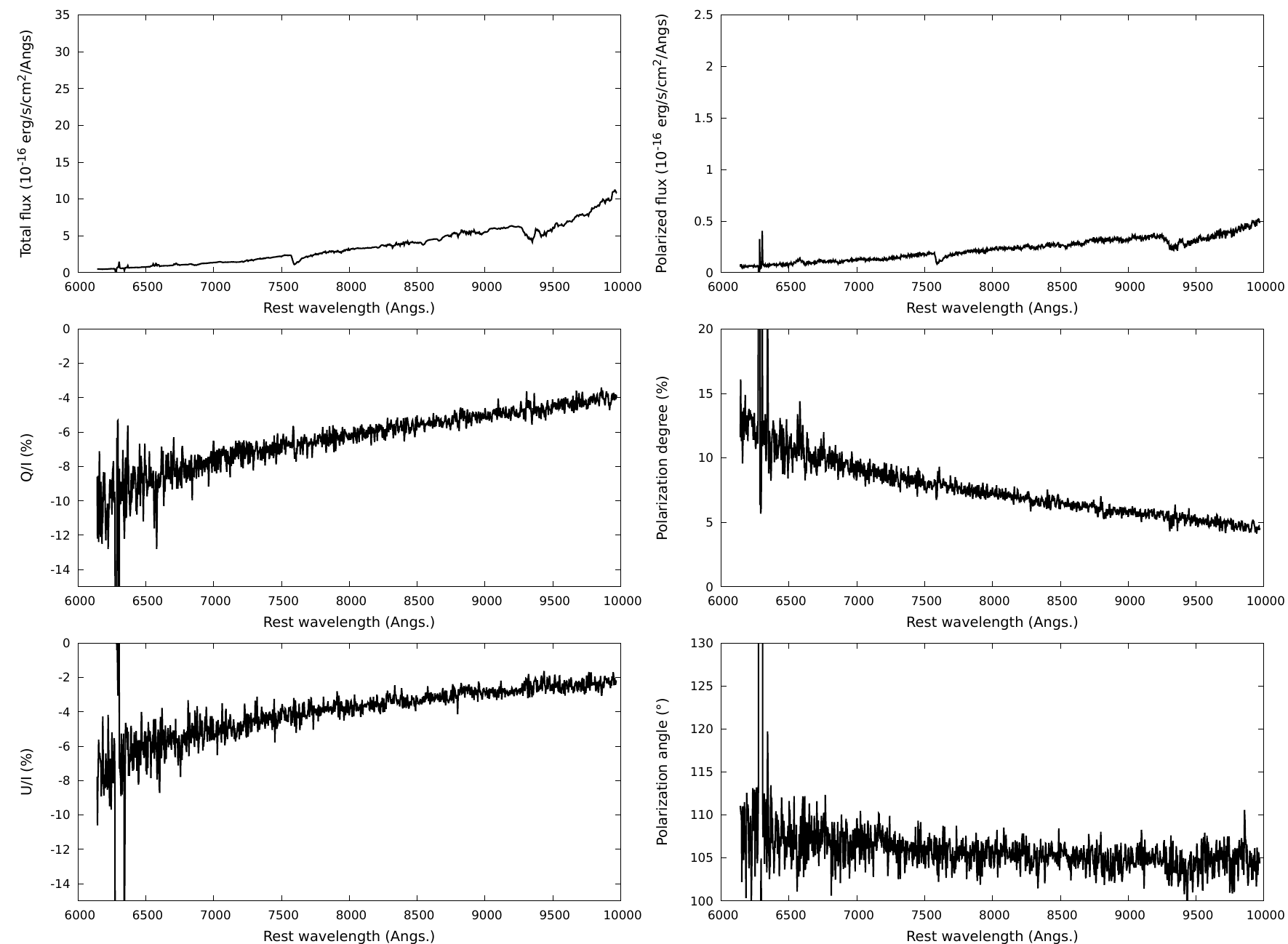}
    \caption{Same as Fig.~\ref{fig:Polarized_knot} but for the region M pinpointed in Fig.~\ref{fig:ADU} and situated between the polarized knot and the SE cloud. The $y$ axes share the same scales as in Figs.~\ref{fig:Polarized_knot} and \ref{fig:SE_cloud} to facilitate comparisons.}
    \label{fig:region_M}
\end{figure*}

We finish our spectropolarimetric analysis of the nuclear region of Cen~A by exploring the total and polarized fluxes of the region M, physically located between the polarized knot and the SE cloud. The extracted spectra are presented in Fig.~\ref{fig:region_M}. Similarly to the SE cloud, the total flux spectrum only shows a heavily reddened continuum locally affected by telluric and stellar absorption lines. However, region M is fainter by a factor 2.2 at 7000~\AA\, in comparison to the polarized knot and SE cloud (who share the same total flux at this specific wavelength). The spectrum is also slightly redder than the SE cloud. Both those arguments imply that region M is undergoing more dust extinction than the SE cloud. If correct, one should expect higher polarization degree values, since polarization induced by dust increases as 2.75\% $\pm$ 0.30\%~ mag$^{-1}$ \citep{Hough1987,Schreier1996}. This is exactly what we find : $P$ is almost as high as in the case of the polarized knot, with a similar decreasing trend from the optical to the near-infrared. Its spectral slope (k = -2.2), similar to the spectral slope of the SE cloud, indicates that both are producing a polarized linear continuum through dichroic absorption but region M is more obscured. This is confirmed again when looking at the polarization angle, which is almost wavelength-independent and very close to the polarization angle of the dust lane. This consistency indicates that the region M and SE cloud polarizations result directly from dust dichroic absorption (i.e. interstellar polarization). 

Now that we have identified a region (region M) whose polarization signal is purely dominated by interstellar polarization and who suffers comparable extinction to the polarized knot (as indicated by the similar polarization levels and spectral slopes), we have a very good tool in hand to extract the intrinsic polarization from the AGN component in the polarized knot spectra.

\subsection{Extracting the AGN component}
\label{Analysis:AGN}

The intrinsic polarization of Cen~A, polluted by interstellar polarization resulting from dichroic absorption, can be retrieved by vector subtraction on the normalized stokes parameter $q$-$u$ plane: 
\begin{equation}
    \begin{split}
        q_{int} = q_{obs} - q_{ISP} \\
        u_{int} = u_{obs} - u_{ISP}
    \end{split}
\end{equation}
where the subscript 'int', 'obs' and 'ISP' stands for the Stokes parameter for the intrinsic (AGN) polarization, observed (AGN + starlight + ISP) polarization and interstellar polarization, respectively. As explained in the previous section, we can reliably use the normalized Stokes parameter $q$ and $u$ of region M to correct the polarization of the Schreier's knot from interstellar polarization in order to retrieve the intrinsic (AGN) polarization. Beware though, this correction will only remove the component due to interstellar polarization. It will not correct the total flux spectrum from the host stellar light, which will have a diluting impact on the intrinsic $P$ (but not on $\theta$) since starlight is essentially unpolarized.

\begin{figure*}
    \centering
    \includegraphics[trim={0cm 0cm 0cm 0cm}, clip, width=\linewidth]{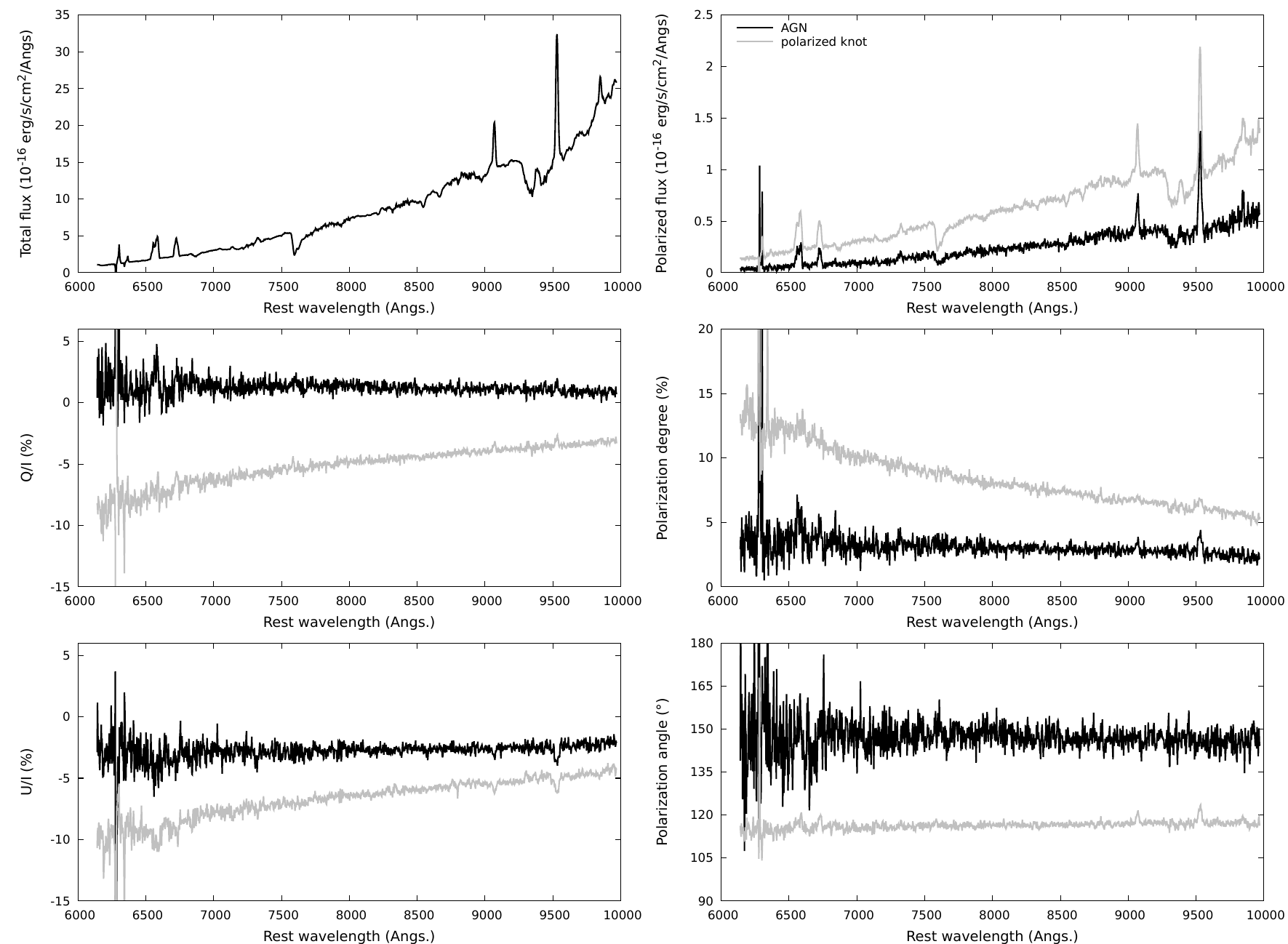}
    \caption{Intrinsic AGN polarization (in black) retrieved from the polarized knot (in grey), corrected for the polluting contribution from the host starlight polarization estimated thanks to region M. All details are similar to Figs.~\ref{fig:Polarized_knot}, \ref{fig:SE_cloud} and \ref{fig:region_M}.}
    \label{fig:Pol_corrected}
\end{figure*}

We show in Fig.~\ref{fig:Pol_corrected} the intrinsic AGN polarization scattered by the polarized knot. Both Q/I and U/I were heavily affected by the correction, leading to a polarization degree that is now much lower than for the polarized knot alone (for which we recall the spectra in gray on the figure), decreasing from $\sim$ 4\% in the optical band to 2\% in the near-infrared. Such wavelength-dependence of $P$ could indicate that the particles responsible for scattering inside the NLR are predominantly dust grains, a hypothesis that is strengthened by the fact that the polarized spectrum is bluer than the total flux, but one should keep in mind the wavelength-dependent, diluting impact of the host starlight. 

We notice that the emission lines are still present in the polarized spectrum, with polarization slightly stronger in the lines than in the neighboring continuum (see Tab.~\ref{tab:line_polarization_AGN}). In typical type-2 AGNs, the narrow lines are weakly polarized ($\sim$ 1\%) and the polarization is thought to originate from dust dichroic absorption. Here, it indicates that both the continuum and the narrow lines are scattered by the same medium, likely a distant portion of the NLR (on the scale of 10 -- 100~pc). We will propose a possible explanation for the unusually high (3 -- 5 \%) polarization degree of the narrow emission lines in the discussion section. Regarding the broad emission lines, we still do not detect any in polarized flux, which tends to make us believe that perhaps there simply is no BLR. 

Finally, \textit{after correcting for the interstellar polarization}, the intrinsic polarization angle has rotated from 115 -- 120$^\circ$ to an almost wavelength-independent value of 146$^\circ$, precisely the value expected for orthogonality with the radio jet position angle. This brings a solid and final conclusion to the interpretation that the nuclear light is scattered by the NLR into our line-of-sight, as seen in typical type2 AGNs. 

\begin{table}
    \caption{Measurement of the intrinsic (AGN) polarization.}     
    \label{tab:line_polarization_AGN}      
    \centering                                      
    \begin{tabular}{c c c c c}        
    \hline\hline              
    Element & Q/I & U/I & $P$ (\%) & $\theta$ ($^\circ$) \\  
    \hline                        
     H$\alpha$+[N~II]$_{\rm cb}$   	&	0.006	&	-0.041	&	4.12	&	  139.3 \\\relax
     H$\alpha$+[N~II]      	&	0.030	&	-0.040	&	4.99	&	  153.5 \\\relax
     H$\alpha$+[N~II]$_{\rm cr}$   	&	0.001	&	-0.042	&	4.25	&	  135.5 \\\relax
     [S~II]$_{\rm cb}$      	&	0.005	&	-0.037	&	3.71	&	  138.3 \\\relax
     [S~II]         	&	0.024	&	-0.039	&	4.58	&	  150.5 \\\relax
     [S~II]$_{\rm cr}$      	&	0.011	&	-0.025	&	2.75	&	  146.7 \\\relax
     [S~III]$_{\rm cb}$     	&	0.011	&	-0.026	&	2.81	&	  146.0 \\\relax
     [S~III]        	&	0.015	&	-0.032	&	3.56  &	  147.7 \\\relax
     [S~III]$_{\rm cr}$    	&	0.009	&	-0.026	&	2.73	&	  144.6 \\\relax
     [S~III]$_{\rm cb}$     	&	0.008	&	-0.026	&	2.70	&	  143.8 \\\relax
     [S~III]        	&	0.017	&	-0.037	&	4.04	&	  147.2 \\\relax
     [S~III]$_{\rm cr}$     	&	0.009	&	-0.024	&	2.56	&	  154.4 \\
    \hline                                          
    \end{tabular}
    \tablefoot{Measurement in the brightest emission lines and in the adjacent 5 spectral bins continuum (cb =  blue continuum, cr = red continuum). Errors are about 0.6\% on $P$ and 4$^\circ$ on $\theta$.}
\end{table}

\subsection{Constraints on the broad line properties}
\label{Analysis:broad_line}

Are we sure the polarized flux plot of Fig.~\ref{fig:Pol_corrected} doesn't show a type-1 spectrum in reflection? To assess the detectability of a potential broad H$\alpha$ line, we evaluate what fraction of the actual broad H$\alpha$ profiles in a representative sample of type-1 AGNs can be confidently detected if it were present in our polarized flux spectrum. The sample consists of 2577 type-1 AGNs and is presented in Jiang et al. (in prep.). This sample is selected from all cataloged SDSS quasars with more than one observation, which we assume is representative of classified type-1 AGNs. We only consider profiles with adequate signal-to-noise ratio (SNR) to define them accurately and superimpose these profiles with proper normalization (their equivalent widths), onto our intrinsic polarized flux spectrum. 

Two model-independent profile parameters: the line centroid $\lambda_{0}$ and line dispersion $\sigma$ are defined as follow.
\begin{equation}
\lambda_{0} = \int f(\lambda)d\lambda
\end{equation}
\begin{equation}
\sigma^{2} = \int f(\lambda)(\lambda-\lambda_{0})^{2}d\lambda
\end{equation}
A detection of a broad H$\alpha$ emission line in our polarized flux — after adding the actual profile — is considered successful if the SNR exceeds 3 at a wavelength that is 1 $\sigma$ redward or blueward from the centroid, effectively indicating detection of the red or blue wing of the profile. An alternative criterion includes identifying excess broad H$\alpha$ flux in the polarized flux spectrum after adding the profile. However, possible contamination from the unresolved narrow H$\alpha$ and [N~II] lines might complicate interpretation. To estimate the noise level in the polarized flux spectrum, we mask out all spectral regions with obvious absorption and emission lines and leverage the fluctuation in the neighboring continuum pixels. This approach yields an average continuum polarized flux SNR of 6 around the H$\alpha$ blend. In examining the 2577 representative type-1 broad H$\alpha$ regions of the comparison sample, we find that more than 99\% of the broad components would be conspicuous if added to our polarized flux spectrum.

We acknowledge that in rare instances, such as the cases of Cygnus~A \citep{Ogle1997} and NGC~3147 \citep{Bianchi2019}, the broad H$\alpha$ line might be more challenging to detect due to extreme broadening (up to 26,000 km/s). However, even considering these extreme cases, the absence of a detectable broad H$\alpha$ line in Cen~A leads us to conclude with $\ge$ 99\% probability that no such broad component exists in our polarized flux spectrum. This strongly suggests that the polarized flux observed in Cen~A is reflected light from the nucleus, but it does not originate from a hidden type-1 AGN as typically expected in the standard Unified Model.

\section{Discussion}
\label{Discussion}

\subsection{Cen~A, a hidden-NLR AGN}
\label{Discussion:hNLR}

As demonstrated in Sect.~\ref{Analysis:AGN}, narrow emission lines are clearly observed in the polarized flux spectra of the knot after subtraction of the contaminating interstellar polarization, with $P$ ranging from 3 to 5\% (including the underlying continuum and not corrected for starlight, so the true $P$ is much higher) and with a polarization position angle exactly perpendicular to the jet axis (see Tab.~\ref{tab:line_polarization_AGN}). The adjacent continuum shows lower (2 -- 3\%, neither corrected for starlight) polarization with an electric vector position angle different by about 10$^\circ$. The most efficient method to achieve such unusual AGN polarization properties is to consider that the base of the NLR is obscured by the torus. Similar to the scenario involving hidden BLRs in type-2 AGNs \citep{Antonucci1993}, the bright innermost part of the NLR may become visible only through perpendicular scattering off the polar winds, naturally producing high polarization degrees and polarization angles perpendicular to the radio structure. The facts that the continuum polarization around the line is lower by a few percent and different by a few degrees in $\theta$ strengthen this scenario : the continuum flux, arising from the vicinity of the supermassive black hole, is already polarized (either by scattering on the accretion flow atmosphere or because it originates from synchrotron emission) and thus, when it scatters onto the outer (unobscured) portion of the NLR towards our line-of-sight, multiple scattering will decrease $P$ and modify $\theta$. The narrow lines emitted inside the obscured portion of the NLR wind, which are unpolarized at emission, will only undergo a single scattering event on average (the NLR being optically thin), maximizing $P$ and fixing $\theta$ to a value orthogonal to the radio axis. 

Now, how extended in the vertical (polar) direction must be the torus in Cen~A? This entirely depends on the exact location of the supermassive black hole. If we rely on the radio map obtained by, e.g., \citet{Ojha2010} or \citet{Janssen2021}, the polarized knot is about 1.4" (26~pc in projected distance) away from the radio core. If we rely on the infrared maps obtained by \citet{Schreier1996}, the projected distance between the AGN core and the knot varies from 0.6" (10.5 pc) in the K band to 2.4" (44.7~pc) in the J band. It implies that the dusty torus should have a vertical size measured from the equatorial plane of the circumnuclear obscurer of 10.5 -- 44.7\footnote{One can roughly estimate the size of the torus in the radial direction knowing that the torus should maintain an aspect ratio of height/radius $\ge$ 0.5 throughout most of the AGN activity phase \citep{Honig2007}. It means that the torus in Cen~A is likely to have an outer radius larger than $\sim$ 90~pc, in agreement with the constraints derived by \citet{Bryant1999}, i.e. 120~pc $\pm$ 20~pc.}~pc. Those half-widths are in very good agreement with the size required to hide the bright, innermost part of the NLR. Indeed, \citet{Balmaverde2016} estimated the radial size of the innermost NLR in a sample of low-luminosity AGNs (LINERs and Seyfert galaxies) observed by the Space Telescope Imaging Spectrograph (STIS) installed on the Hubble Space Telescope (HST) and found that the bright, innermost region of the NLR usually extend up to 10~pc in these sources. It means that, with a torus extending up to 10.5 -- 44.7~pc in the polar direction, the innermost NLR would be obscured and its signal could only be seen thanks to orthogonal scattering off the dust grains in the outermost NLR. Interestingly, the idea of a vertically extended torus in Cen~A was already presented by \citet{Bryant1999} but using a completely different methodology and argumentation, based on J-, H- and K-band images and 0.9 -- 2.5 microns spectra of the nuclear regions of Cen~A. In their paper, the authors successfully reproduced the extinction map obtained with the J-K image using a circumnuclear torus perpendicular to the radio jet and sustaining a total vertical thickness of 75~pc $\pm$ 4~pc, so a vertical extension of about 37.5~pc from the equatorial plane, sensibly in-line with our findings. 

An immediate consequence of such a vertically extended torus should be the detection of mid-infrared photons within several (tens of) parsecs of the radio core. \citet{Radomski2008} have shown that most of the 8.8~$\mu$m and 18.3~$\mu$m Cen~A emission comes from a compact spherical core of size $\le$ 5~pc, but there is a large amount of diffuse mid-infrared emission around this core, extending to more than ten parsecs. If, as postulated by \citet{Shi2006}, the torus is not a monolithic structure but rather a multilayered system with dust density and temperature decreasing as we move away from the core, the compact mid-infrared dust emission detected by \citet{Radomski2008} would be the inner, hotter part of it, while the outer ($\ge$ 10~pc), colder layer, consisting of gas and dust, should be brighter in the far-infrared continuum and CO/H2 emission instead. This was indeed detected in several AGNs thanks to ALMA observations \citep{Combes2019,Garcia2019,Garcia2021,Garcia2024}, and more precisely in Cen~A by \citet{McCoy2017}, who have shown that the 1.3~mm and 3~mm continuum emission is isotropic and spans about 110~pc in size. All this reinforces our conclusions about the unusual size of the torus in Cen~A, accounting for the density gradient proposed by \citet{Shi2006}.

This and the polarization properties of Cen~A makes it a truly exotic type of AGN, but it is not unique. In fact, this is the third case where highly polarized narrow lines, perpendicular to the jet, are reported. The two other cases are the highly inclined Seyfert 1.9 galaxy NGC~4258 \citep{Barth1999}, whose position angle of polarization is oriented nearly parallel to the projected plane of the masing disk (86$^\circ$, \citealt{Miyoshi1995,Wilkes1995}) and the FR~I radio galaxy 3C~270 (also known as NGC~4261, Jiang et al., in prep.), whose giant torus is famously directly visible as absorption against the bulge starlight \citep{Jaffe1993}. The case of NGC~4258 is particularly interesting since the authors also found narrow lines polarized at 5 -- 6\% in general, with a linear continuum polarization of 0.3 -- 0.4\% that has a polarization angle different by 10$^\circ$ with respect to the emission line $\theta$. In addition, there is no evidence for the presence of broad H$\alpha$ lines in the polarized flux spectrum in any three of these objects. The resemblance with Cen~A is striking. Finally, we note that these objects, showing LINER-like line ratios, lie at the low bolometric luminosity end of both radio-loud and radio-quiet AGNs. These active nuclei probably manifest most of their radiative power through synchrotron emission \citep{Younes2012}.

The requirement of an unusually large vertical scale for the obscuring torus to explain the high polarization of the narrow lines sets these AGNs apart from the majority. Hence, we deem it necessary to classify these AGNs as "hidden-NLR AGNs" with giant obscuring tori (or any geometry of the circumnuclear region, such as, e.g., a flared disk) hiding the NLR. Although rarer than tori not exceeding 1~pc in vertical size \citep{Hopkins2012}, circumnuclear structures extending more than 10~pc in height are entirely plausible solutions of multiscale smoothed particle hydrodynamic simulations \citep{Hopkins2010}. This would inevitably result in high covering factors.

\subsection{\textbf{The nature of Cen~A's observed polarization}}
\label{Discussion:POlarization}

More than half a century ago, early Cen~A polarimetric observations achieved by \citet{Elvius1964} already revealed up to 6\% polarization in a 27-arcsecond aperture, with angles parallel to the dark lane, starting a debate about the nature of the observed polarization in Cen~A. If the author attributed it to either interstellar absorption or optical synchrotron emission without a definitive conclusion, soon two main hypotheses emerged.

One supported a synchrotron nature for the observed polarization. \citet{Bailey1986} detected high K-band polarization with angles nearly perpendicular to the radio jet, suggesting synchrotron emission dominates in the near-infrared due to reduced dust extinction in comparison to the optical band. They proposed that Cen~A behaves like a "mis-directed BL Lac object", so that the observed polarization is purely from synchrotron emission.

The other hypothesis supported dust scattering. \citet{Packham1996} observed increasing polarization degree with smaller apertures and wavelength in J, H, and Kn bands, with Kn-band polarization perpendicular to the jet. Millimetric observations at 800 and 1100 $\mu$m were found to be consistent with the nucleus of Cen~A being unpolarized at these wavelengths. They attributed those signatures to polar scattering rather than synchrotron emission, noting that polarization due to scattering is uncommon in FR I radio galaxies \citep{Capetti2007}, but more typical in high ionization radio galaxies.

In light of our work, we can safely conclude that the polarization in Cen~A results from a combination of dust scattering and interstellar polarization from dichroic absorption, which adequately explains the observed phenomena in the optical, near-infrared and millimeter bands. The last question that remains unanswered is: what is the origin of the continuum scattered by dust in the NLR? Thermal emission or beamed synchrotron emission?

\subsection{The nature of this FR~I prototype central engine}
\label{Discussion:Engine}

We now consider the energy source for the warm dust and narrow-line radiation. To do so, we explore three arguments, listed in ascending order of energy.

The infrared arguments -- The near-infrared spectrum of Cen~A lacks any region consistent with $F_\nu \propto \nu^2$, ruling out Rayleigh-Jeans stellar emission, despite the detection of calcium triplet absorption \citep{Simpson1998,Bryant1999}. This suggests that the central engine's emission is heavily absorbed, hiding potential broad lines and highlights the importance of spectropolarimetry to probe this environment. While dust emission near the sublimation temperature could contribute, the kilo-parsec-scale dust lane, which becomes nearly transparent at 2 microns, is a plausible alternative explanation. This is supported by the sharp flux rise and 17\% starlight-corrected polarization at 2 microns, with a position angle perpendicular to the jet \citep{Bailey1986}, indicating scattered light near the nucleus. Mid-infrared observations further support a non-thermal origin. The inferred dust bolometric luminosity, $\sim$ 4.94 $\times$ 10$^{41}$ erg s$^{-1}$, suggests heating by ultraviolet and X-rays rather than multi-temperature blackbody radiation from an accretion flow. Thermal emission from such a flow would produce a characteristic continuum spectrum, including strong ultraviolet-to-optical features and a Rayleigh-Jeans tail, none of which are observed. Instead, the spectral shape, combined with the polarization and flux properties, points to synchrotron radiation as the dominant mechanism in the near-infrared and mid-infrared, with some dust heating contributing secondarily.

The ultraviolet arguments -- Ultraviolet heating of the innermost AGN dust requires powerful mechanisms, such as thermal emission from accretion or synchrotron radiation. If the central engine of Cen~A possesses a standard accretion flow and/or a standard broad line region (BLR -- that could also be responsible for the continuum emission according to \citealt{Netzer2022}), it should exhibit the characteristic "big blue bump" (BBB), the thermal emission peaking in the ultraviolet. However, previous studies, including \citet{Marconi2001}, \citet{Rieger2009}, and \citet{Marin2023}, reported the absence of a BBB in the optical and ultraviolet spectra of Cen~A, based on archival data and prior observations. This absence deviates from the expectations of standard accretion models, suggesting an alternative mechanism for the observed continuum. The need for ultraviolet heating is, however, confirmed by independent observations. Using the Infrared Spectrograph on the Spitzer Space Telescope, \citet{Quillen2008} detected high-ionization lines such as [Si IV] and [Ne V], requiring sufficient ionizing photons in the continuum. While we do not directly measure the BBB in our spectropolarimetric data, its absence remains consistent with prior findings. A plausible source of the required ionizing continuum could be beamed synchrotron emission, which compensates for the lack of a BBB, as suggested by models such as \citet{Rodriguez2020}. Alternatively, it is possible that earlier epochs featured a detectable BBB/BLR signature when the incident continuum illuminated the high-ionization NLR clouds observed in the mid-infrared, but that this component has since diminished, suggesting variability in Cen~A.

The X-rays arguments -- The X-ray properties of Cen~A provide essential insights into the nuclear region and its obscuring structures. The unabsorbed X-ray luminosity in the 2–10 keV band indicates an Advection-Dominated Accretion Flow (ADAF), consistent with Cen~A’s low Eddington ratio \citep{Evans2004}. This supports the absence of a BLR, aligning with our spectropolarimetric non-detection of broad H$\alpha$, as ADAF models predict \citep{Liu2009}. The X-ray polarization measured by IXPE \citep{Ehlert2022}, although limited to an upper bound of $P \le$ 6.5\%, preventing direct determination of the polarization angle, supports synchrotron self-Compton (SSC) emission, consistent with the spectral shape and polarization geometry from radio to X-rays \citep{Marin2023}. This is reinforced by our optical-near-infrared polarized spectrum, which also suggests synchrotron origins for the central emission. Complications arise from the narrow Fe~K$\alpha$ line, which requires quasi-isotropic X-ray emission to irradiate the Compton-thin absorber \citep{Tzanavaris2021}. While synchrotron-dominated X-rays would be anisotropic, our spectropolarimetric data suggest a two-component model: quasi-isotropic X-rays from the ADAF interacting with the torus, and beamed synchrotron radiation contributing to scattered light from the inner narrow-line region. This is supported by X-ray reverberation mapping \citep{Iwata2024}, showing distinct components at $\sim$0.2 pc and $\ge$1.7 pc.

\begin{figure}
    \centering
    \includegraphics[trim={0cm 0cm 0cm 0cm}, clip, width=\columnwidth]{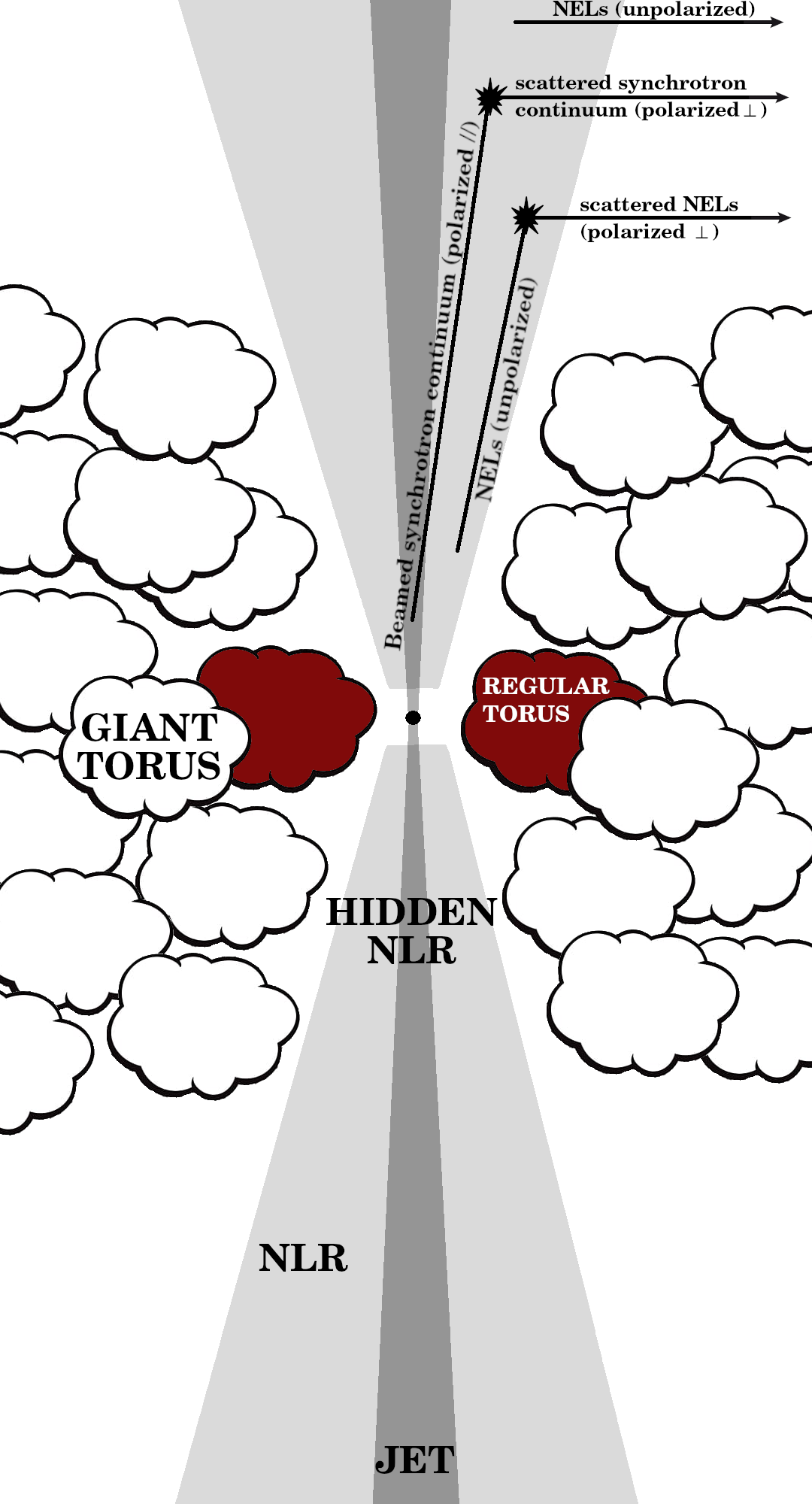}
    \caption{Schematic view of a hidden-NLR AGN. The black dot at the center represents the supermassive black hole, in the vicinity of which jets (in dark grey) are produced. The jet passes through the NLR (in light grey) and interacts with it along its path. The circumnuclear material, labeled as "torus" for brevity, is shown using clouds. In the case of a regular AGN, the typical optical size of the torus is represented in red, while in the case of a hidden-NLR AGN, the giant torus (in white, $\ge$ 10~pc) occupies a much larger volume, effectively hiding the innermost part of the NLR. We show on the figure how the unpolarized narrow emission lines (NELs) from the hidden NLR become polarized perpendicular to the jet radio axis due to scattering onto the NLR outermost, unobscured part. We also show how the beamed synchrotron continuum emission, likely polarized parallel to the jet axis at origin, is scattered in our line-of-sight and becomes perpendicular to the jet axis (although with a $\sim$ 10$^\circ$ difference due to the fact that the continuum was already polarized before this interaction). Finally, we show that regular, unpolarized, low-ionization NELs are still produced and observed in the outermost part of the NLR.}
    \label{fig:Schema}
\end{figure}

In conclusion, given the absence of a big blue bump in the spectral energy distribution of Cen~A, the absence of broad lines in polarized flux, the low luminosity of the continuum source, the high energy constraints brought by X-ray polarimetry and the near-infrared polarization properties of the source, we conclude that the continuum source of photons in Cen~A is from beamed synchrotron emission that is scattered towards us by material in the polar direction of the AGN, beyond the vertical limits imposed by a torus more voluminous than usual. This is a \textit{significant} and \textit{unprecedented} discovery as it confirms a prediction made by \citet{Blandford1978} and, later, by \citet{Antonucci1990} and \citet{Wills1992}, who stated that "beamed blazar radiation [...] can in principle sometimes be seen in reflection from Earth" \citep{Antonucci1990}. A schematic view of a hidden-NLR AGN is presented in Fig.~\ref{fig:Schema} to better visualize the geometry and physics involved in such atypic AGNs.

\section{Conclusion}
\label{Conclusion}

We have obtained spectropolarimetric data of the nearest radio galaxy, Cen~A (Centaurus~A) using the VLT/FORS2. We presented 6152 -- 9988~\AA\ total and polarized fluxes spectra with an effective resolving power of around 1000, for several regions nearby the (dust obscured) location of the AGN. A direct measurement of the AGN continuum and line fluxes is not achievable due to the presence of a kilo-parsec dust lane bisecting the field-of-view, effectively hiding the AGN. Focusing on three different regions within the slit, namely a polarized knot already revealed by \citet{Schreier1996}, a bright and blue cloud south-east of the knot, and the region in-between the two aforementioned targets, we discovered that only the polarized knot is showing strong and narrow emission lines ascribed with AGN activity. The SE cloud appears to be pure starlight with low foreground extinction and the third region is also starlight but suffering similar extinction as the polarized knot. Taking advantage of this fact, we corrected the polarized knot signal from dichroic absorption and retrieved the intrinsic polarization arising from AGN light that has scattered on the cloud. We found a starlight-diluted polarization degree of 2 -- 4\%, decreasing from the optical to the near-infrared, potentially indicating that Mie scattering is the dominant polarizing mechanism in the knot, which is likely part of the extended NLR of the AGN. More importantly, the intrinsic polarization position angle is found to be almost wavelength-independent and precisely perpendicular to the radio jet axis. Our findings show beyond doubt that the light is scattered into our line of sight, consistent with 2-micron polarimetric observations where extinction is much less significant.

Several crucial findings emerged from our interstellar polarization corrected spectrum. First, we detected no polarized broad lines, neither in the total nor in the polarized fluxes. A stringent upper limit on the equivalent width of the broad H$_{\alpha}$ line to be seen in the polarized flux given various FWHM profiles was derived and we concluded that there is no broad lines in Cen~A with 99\% probability. Second, we found that the narrow emission lines are polarized, stronger than that of the continuum, with a polarization angle exactly perpendicular to the radio axis of the jets, while the continuum has a polarization angle deviating by 10$^\circ$ with respect to orthogonality. 

To explain such an unusual set of properties, we suggest that Cen~A defines a very exotic class of "hidden-NLR AGNs", where the innermost and brightest part of the NLR is obscured by a circumnuclear torus whose vertical extension extends for several tens of parsecs. The synchrotron continuum, most likely intrinsically polarized parallel to the jet axis \citep{Perlman1999,Park2022}, irradiates the NLR, which in turn produces unpolarized narrow emission lines. Both the continuum and the narrow lines must travel through the dust funnel and can only escape by scattering onto the uppermost part of the NLR. The narrow emission lines gain a relatively large polarization degree (a few percent) and a polarization position angle exactly perpendicular to the radio jet, while the synchrotron continuum sees its previous polarization altered by this orthogonal diffusion, resulting in lower $P$ and a value of $\theta$ close, but slightly deviating, from perpendicularity. The concept is very similar to regular type-2 AGNs, where the broad lines, obscured by the dusty torus, are only revealed in polarized light; a key point in common with the unification model is the perpendicular polarization angle due to polar scattering.

We concluded our analysis of Cen~A by deducing that the continuum source of photons is very likely synchrotron emission, based on several key observations: the absence of the Rayleigh-Jeans region in the near-infrared spectrum, the lack of a BBB in the ultraviolet spectrum and the consistency of X-ray polarization data with synchrotron self-Compton models. These findings collectively indicate that synchrotron emission, rather than thermal emission from an accretion flow, is the dominant mechanism for the observed continuum, that is subsequently scattered toward us in the outermost part of the NLR. This discovery is both significant and unprecedented, as it confirms a prediction initially made by \citet{Blandford1978} which stated that beamed synchrotron jets should be observed in reflection in certain very specific cases.

The next step would be to survey the population of AGNs to discover other examples of hidden-NLR AGNs. At least two other sources have been confirmed so far (NGC~4258 and 3C~270) but, to look for such objects, one shall concentrate on misdirected BL~Lac AGNs, i.e. low power non-thermal radio galaxies with a jet oriented close to our line-of-sight (but not of the class of blazars). Their spectra should just show LINER lines and a continuum, with no broad lines in total flux. Their polarized flux should reveal a complete absence of broad emission lines, a moderately polarized continuum and highly polarized narrow emission lines, polarized perpendicular to the radio jet position angle.

\begin{acknowledgements}We would like to thank the anonymous referee for his/her constructive comments that helped to increase the quality of this paper. Based on observations collected at the European Southern Observatory under ESO programmes 111.24ZY.001, 111.24ZY.002, 111.24ZY.003. F.M. would like to acknowledge the support of the CNRS, the University of Strasbourg, the PNHE and PNCG. This work was supported by the "Programme National des Hautes Énergies" (PNHE) and the "Programme National de Cosmologie et Galaxies (PNCG)" of CNRS/INSU co-funded by CNRS/IN2P3, CNRS/INP, CEA and CNES. DH is research director at the F.R.S-FNRS, Belgium. CRA acknowledges support from the Agencia Estatal de Investigaci\'onof the Ministerio de Ciencia, Innovaci\'on y Universidades (MCIU/AEI) under the grant ``Tracking active galactic nuclei feedback from parsec to kiloparsec scales'', with reference PID2022$-$141105NB$-$I00 and the European Regional Development Fund (ERDF).
\end{acknowledgements}

\bibliographystyle{aa} 
\bibliography{bibliography} 

\end{document}